\newcommand{\ndmap}{Ni(C$_5$H$_{14}$N$_2$)$_2$N$_3$(PF$_6$)}
\newcommand{\ndmaz}{Ni(C$_5$H$_{14}$N$_2$)$_2$N$_3$(ClO$_4$)}
\newcommand{\nenp}{Ni(C$_2$H$_8$N$_2$)$_2$NO$_2$(ClO$_4$)}

\documentstyle[prl,aps,twocolumn]{revtex}
\begin{document}
\input{psfig.sty}
\draft

\twocolumn[\hsize\textwidth\columnwidth\hsize\csname
@twocolumnfalse\endcsname

\title{Field-induced commensurate long-range order in the Haldane-gap system \ndmaz.}

\author{A. Zheludev}
\address{Physics Department, Brookhaven National Laboratory,
Upton, NY 11973, USA.}

\author{Z. Honda$^{\ast}$ and K. Katsumata$^{\sharp}$}
\address{RIKEN (The Institute of Physical and Chemical Research),
Wako, Saitama 351-0198, Japan. ($\ast$) Present address: Faculty
of Engineering, Saitama University, Urawa, Saitama 338-8570,
Japan. (${\sharp}$) Present address: RIKEN Harima Institute,
Mikazuki, Sayo, Hyogo 679-5148, Japan.}

\author{R. Feyerherm and K. Prokes}
\address{ Hahn-Meitner-Institut GmbH,
Glienicker Str. 100, 14109 Berlin, Germany.}

\date{\today}
\maketitle
\begin{abstract}
High-field neutron diffraction studies of the new
quantum-disordered $S=1$ linear-chain antiferromagnet \ndmaz\
(NDMAZ) are reported. At $T=70$~mK, at a critical field
$H_c=13.4$~T applied along the (013) direction, a phase transition
to a commensurate N\'{e}el-like ordered state is observed. The
results are discussed in the context of existing theories of
quantum phase transitions in Haldane-gap antiferromagnets, and in
comparions with previous studies of the related system \ndmap\
(NDMAP).
\end{abstract}

 ]

\section*{Introduction}
One of the most exciting recent developments in experimental
studies of one-dimensional (1D) magnets was the observation of a
field-induced quantum phase transitions in two new quasi-1D $S=1$
antiferromagnets (AFs) \ndmaz\ (NDMAZ) \cite{Honda97} and \ndmap\
(NDMAP) \cite{Honda98,Chen01}. In the absence of an external
magnetic field the ground state of such systems is a non-magnetic
singlet. Long-range magnetic order is totally destroyed by quantum
spin fluctuations even at $T=0$: spin correlations are short
range, and decay exponentially on a length scale of a few lattice
repeats \cite{Haldane83}. The main feature of the magnetic
excitation spectrum is the so-called Haldane energy gap $\Delta$
\cite{Haldane83}. Such magnetically disordered systems may be
described as a ``quantum spin liquids''. The effect of magnetic
field is to suppress zero-point fluctuations, and restore a
gapless spectrum. The result is a quantum phase transition at a
certain critical field $H_c\sim \Delta/(g\mu_{\rm B})$, to a
N\'{e}el-like state with long-range staggered magnetic order, that
may be characterized as a ``spin solid''
\cite{AffleckH90,AffleckH91,Tsvelik90,Sachdev94,Yajima94,Mitra94}.
The transition is driven by a complete softening of one member of
the Haldane excitation triplet, subject to Zeeman splitting in an
external field, and was shown to be equivalent to Bose
condensation in one dimension \cite{AffleckH91,Nicopoulus91}.
While the high-field transition was predicted theoretically a
while ago, NDMAZ ($\Delta\approx 1.7$~meV) \cite{Honda97,Koda99}
and NDMAP ($\Delta\approx 0.5$~meV)
\cite{Honda98,Honda99,Honda99JAP,Honda00,Honda00JAP,Zheludev01}
were the first systems where it was actually observed and
investigated by means of magnetization, magnetic resonance, and
specific heat measurements. For many ``veteran'' Haldane-gap
materials, such as Y$_2$BaNiO$_5$ ($\Delta\approx 9$~meV
\cite{YBANO}), the critical field values are prohibitively high.
In other well-studied compounds, such as \nenp\ (NENP, $\Delta
\approx 1$~meV \cite{NENP}), the transition does not occur at all,
and is replaced by a broad crossover phenomenon, due to certain
structural features \cite{Chiba91,Kobayashi92,Enderle00}.

A direct evidence of long-range AF order above $H_{c}$  can only
be obtained in neutron diffraction experiments. Recently, such
measurements were performed for NDMAP, and brought valuable
insight in the nature of the transition and the high-field phase
\cite{Chen01}. It was found that long-range ordering can be either
3-dimensional or 2-dimensional, depending on the direction of
applied field, but always commensurate with the underlying crystal
lattice. This study was enabled by the rather small value of the
gap energy and hence the critical field ($H_c\sim 5$~T) in this
material. Performing a similar neutron diffraction investigation
of NDMAZ is crucial to establish the universal nature of the
observed behavior, but is, unfortunately, much more challenging
from the technical point of view. In NDMAZ the Haldane excitation
triplet is split by single-ion anisotropy: $\Delta_x=1.61$~meV,
$\Delta_y=1.76$~meV and $\Delta_z=2.8$~meV for excitations
polarized along the $a$, $b$ and $c$ crystallographic axes,
respectively \cite{Honda97,Koike00}.  The spin chains in NDMAZ are
parallel to the $c$ axis. In the presence of anisotropy the
critical field depends on sample orientation and is given by
$H_c^{\alpha}=\sqrt{\Delta_{\beta}\Delta_{\gamma}}/g\mu_{\rm B}$.
The expected values are $H_c^{\bot}\approx 18$~T for $\bbox{H}\bot
\bbox{c}$ and $H_{c}^{z}\approx 12$~T for $\bbox{H}\| \bbox{c}$.
Even the lower critical field, for $\bbox{H}\| c$, is not easy
(though possible) to achieve in a neutron scattering experiment:
for low-temperature studies the current limit is about 14.5~T.
Sadly, an experimental geometry with $\bbox{H}\| c$ can not be
realized. The softening of the Haldane-gap excitation occurs at
the 1D AF zone-center, i.e., at a momentum transfer $q_{\|}=\pi/c$
along the chain axis. It is at this position that magnetic Bragg
reflections, characteristic of the ordered high-field phase, are
expected to appear. However, the scattering plane accessible when
using a split-coil cryomagnet is perpendicular to the field
direction. To enable a $\pi/c$ transfer along the chains the
sample must thus be mounted with the $c$ axis as closely in the
horizontal plane as possible, which pushed the critical field up,
towards its maximum value of 18~T. In the present paper we report
the first neutron scattering observation of field-induced
long-range ordering in NDMAZ that was achieved by using the most
powerful split-coil magnet available, and by mounting the sample
with the chain axis at a small angle relative to the field
direction  to minimize $H_c$, yet sufficiently tilted to enable
momentum transfers of $\pi/c$ along the chains.

\section*{Experimental procedures}
The crystal structure  of NDMAZ (Figure~\ref{struct}) is
monoclinic, space group $C2$, with lattice constants
$a=18.898$~\AA, $b=8.171$~\AA, $c=6.111$~\AA\ and
$\beta=98.27^{\circ}$ \cite{Yamashita95}. The magnetism is due to
$S=1$ Ni$^{2+}$ ions that are bridged by azido groups and form
distinct chains propagating along the $c$ crystallographic axis.
Unlike in NDMAP, where the Ni-sites form a body-centered lattice
\cite{Monfort96}, in NDMAZ the arrangement of spin carriers is
$c$-face centered. In the present study we used a 0.5~g fully
deuterated sample that was found to contain two single-crystal
twins of roughly equal volume, sharing the $b$ and $c$ axes. The
mosaic spread of each of the two crystallites was found to be
$0.4^{\circ}$.

The measurements were carried out at the E4 2-axis diffractometer
installed at the Hahn-Meitner-Institute,  Germany, using a neutron
beam of wavelength $\lambda=2.44$~\AA. The sample was mounted in a
14.5~T split-coil cryomagnet  with the $[0 1 3]$ real-space axis
vertical, i.e., parallel to the field direction. The angle between
the direction of magnetic field and the chain axis was
$\phi=24^{\circ}$. All measurements were carried out at $T=70$~mK,
achieved by using a $^3$He-$^4$He dilution refrigerator magnet
insert. Nuclear Bragg reflections originating from the two members
of the twinned crystal could be easily identified except along the
$(h,0,0)$ reciprocal-space direction where they exactly overlap.

\section*{Results}
At $T=70$~mK, in magnetic fields exceeding $H_c\approx 13.4$~T new
Bragg reflections were detected in both crystallites at
$(h,\case{2n+1}{2},\case{2m+1}{2})$ ($h$, $m$, $n$-integer)
reciprocal space positions. Figure~\ref{exdata} shows scans across
the $(0,1.5,-0.5)$ reciprocal-space point taken above and below
the critical field. The shaded areas represent experimental
resolution, as measured on the $(4,0,0)$ nuclear Bragg reflection,
that corresponds to almost the same momentum transfer as
$(0,1.5,-0.5)$. Gaussian fits to the data show that the observed
width of the magnetic peaks is resolution-limited in both
directions. The measured field dependence of the $(0,1.5,-0.5)$
peak intensity is shown in Fig.~\ref{vsh}. The solid line
represents a power-law fit to the data, that gives an estimate for
the critical field $H_c=13.4(1)$~T and the order-parameter
critical index $\beta=0.21(2)$.

All together, 5 inequivalent magnetic reflections were observed in
the high-field phase on the $(h,1.5,-0.5)$ reciprocal-space rod.
For NDMAP it was previously possible to measure the absolute
values of magnetic Bragg intensities and deduce the spin structure
in the high-field phase, through a callibration against measured
nuclear peak intensities \cite{Chen01}. For technical reasons such
analysis was impossible in the present experiment. The sample was
aligned at room temperature outside the cryostat. Even if this
alignment was perfect, an additional misalignment of the order
1--2 degrees was unavoidable when inserting the sample into the
well of the cryomagnet and cooling to base temperature. As a
result, the  $(0,1,3)$ reciprocal-space plane  of the sample was
slightly misaligned with the horizontal scattering plane. The
14.5~T magnet (unlike the 9~T magnet used for NDMAP) could not be
tilted to compensate for this effect. Under these circumstances
the absolute intensities of Bragg reflections measured in rocking
curves are unreliable, since some of the peaks can, in fact, be
outside the scattering plane, and are only picked up by the
non-zero vertical angular acceptance of the detector.

\section*{Discussion}
Despite the obvious limitations imposed on the measurements by
technical aspects of the experiment, the results presented above
contain important information pertaining to the nature of the
high-field phase in NDMAZ. The principal result of this work is
that the magnetic Bragg reflections, characteristic of the
high-field state, are detected at strictly commensurate positions.
Similar behavior was previously seen in NDMAP \cite{Chen01}.
Theoretical studies of isolated Haldane spin chains in applied
field suggested a potential for incommensurate ordering
\cite{AffleckH91,Yajima94}. In particular, the longitudinal part
of the real-space correlation function for a single chain at
$H>H_c$ contains a $\cos$-term with a built-in periodicity that
depends on $(H-H_c)$ (Eq.~2.8 in Ref.~\onlinecite{Yajima94}). This
incommensurability is in many ways similar to that in
$S=1/2$-systems in applied fields \cite{Pytte74,Muller81}, and can
be understood from a fermion mapping of the Heisenberg Hamiltonian
\cite{AffleckH90,AffleckH91,Tsvelik90}. The reason why
incommensurate ordering does not actually occur lies in the
power-law prefactor to the $\cos$-term: it decays rapidly with
distance, and the Fourier transform of the spin correlation
function does not contain any singularities or even local maxima
at incommensurate wave vectors \cite{Sachdev94,Zal}.

In NDMAP, depending of field direction, the high-field phase was
found to be ordered in either 2 or 3 dimensions \cite{Chen01}. For
NDMAZ, at least in the geometry of the present experiment, it
appears that above $H_c$ the system is in true 3D AF ordered
state. The observed Bragg reflections are resolution-limited along
the $(1,0,0)$ and $(0,3,-1)$ directions. While it is conceivable
that the peaks can be broadened or even rod-like along the
$[0,1,3]$ direction, such a coincidence seems unlikely. Moreover,
there are intrinsic physical reasons for NDMAZ to prefer a 3D
ordered state at high fields. In NDMAP the signature of 2D
ordering are Bragg rods along the $(1,0,0)$ direction that
represent the absence of static spin correlations along the $a$
axis, while 2D ordering occurs within the $(b,c)$ planes
\cite{Chen01}. This, in turn, is a consequence of extremely weak
magnetic interactions along the 18\AA~ crystallographic $a$ axis,
and substantial inter-chain coupling within each $(b,c)$ plane
\cite{Zheludev01}. The coupling between $(b,c)$ planes in NDMAP is
also highly frustrated, due to the body-centered structure. In
NDMAZ the structure within each plane is quite similar to that in
NDMAP, and similar inter-chain interaction strengths along the $b$
axis can be expected. The crucial difference between the two
structures lies in the interactions between Ni$^{2+}$ sites from
adjacent planes. Indeed, the monoclinic distortion in NDMAZ
partially lifts the geometric frustration of the face-centered
arrangement of the Ni-ions. In addition, the distance between
nearest-neighbor Ni-Ni sites from adjacent planes in NDMAZ is
smaller than in that NDMAP, which favors more potent inter-plane
interactions. Future high-resolution inelastic neutron scattering
studies will be required to fully resolve the issue of inter-chain
coupling in the NDMAZ material.

The measured value $H_c=13.4$~T for $\bbox{H}\|[0 1 3]$ in NDMAZ
is in reasonable agreement with the previously measured Haldane
gap energies \cite{Koike00}. Indeed, for the axially symmetric
case, with a magnetic field applied at an angle $\phi$ to the
anisotropy axis, $\mu_B H_c=\Delta_z\Delta_{\bot}/\sqrt
{g_z^2\Delta_z^2\cos^2\phi+g_{\bot}^2\Delta_z\Delta_{\bot}\sin^2\phi}$
\cite{Chen01}. In this formula $g_z=2.11$ and $g_{\bot}=2.21$ are
components of the Ni$^{2+}$ gyromagnetic tensor along and
perpendicular to the anisotropy axis, respectively \cite{ZH}. For
the geometry realized in the present experiment $\phi=24^{\circ}$
and the formula gives $H_c=13.8$~T. Note that inter-chain
interactions always reduce the gap energy at the 3D AF
zone-center, which should lead to smaller actual critical field
values than given by the above equation.

The observed value of the critical index $\beta=0.21$ for NDMAZ is
similar to that found in NDMAP and is also consistent with
theoretical expectations. For a field applied at an angle to the
anisotropy axis, the direction of ordered moment in the high-field
phase is predetermined. On the one hand, it is the Haldane gap
excitation polarized transverse to the applied field that softens
at $H_c$ and the resulting ordered moment must also be normal to
$\bbox{H}$, {\it i. e.}, be confined to the $(0,1,3)$ plane. On
the other hand, it can be expected to lie in the magnetic easy
plane $(0,0,1)$. The moment is thus necessarily aligned along the
$a$ axis, and we are dealing with an Ising-type transition. In
fact, for an isolated chain, the transition  at $H_c$ is expected
to fall in the 1+1-dimensional Ising universality class, with $H$
taking the role of temperature \cite{AffleckH91} and
$\beta=0.125$. For NDMAZ however, inter-chain interactions along
the $b$ axis can be considered substantial, and 3D Ising behavior,
with $\beta\approx 0.31$, may be expected. The measured critical
exponent falls in between these two values and may be a signature
of a dimensional crossover regime.

\section*{Conclusion}
NDMAZ is the second Haldane-gap system where field-induced
commensurate antiferromagnetic ordering is observed by means of
neutron diffraction. The similarity of the behaviour observed in
NDMAZ and NDMAP suggests that it represents a generic trend for
weakly coupled $S=1$ quantum spin chains.

\acknowledgements

We would like to thank S. M. Shapiro (BNL) for insightful
discussions, and P. Smeibidl and S. Kausche (HMI) for technical
assistance. Work at Brookhaven National Laboratory was carried out
under Contract No. DE-AC02-98CH10886, Division of Material
Science, U.S.\ Department of Energy. Work at RIKEN was supported
in part by a Grant-in-Aid for Scientific Research from the
Japanese Ministry of Education, Culture, Sports, Science and
Technology


\begin{figure}
\caption{A schematic view of the antiferromagnetic spin chains in
the NDMAZ crystal structure. Only NiN$_6$ octahedra (grey) and
N-atoms (black) are show.} \label{struct}
\end{figure}

\begin{figure}
\caption{$h$- (left) and $k$-scans (right) across the
$(0,1.5,-0.5)$ reciprocal-space point in a deuterated NDMAZ sample
at $T=70$~mK for magnetic fields above (solid circles) and below
(open circles) the critical field $H_c=13.4$~T. The field is
applied along the $[0,1,3]$ direction. Shaded areas represent the
measured experimental wave vector resolution.} \label{exdata}
\end{figure}
\begin{figure}
\caption{Measured field dependence of the $(0,1.5,-0.5)$ magnetic
Bragg peak intensity in NDMAZ at $T=70$~mK. The solid line is a
power-law fit to the data, as described in the text.} \label{vsh}
\end{figure}

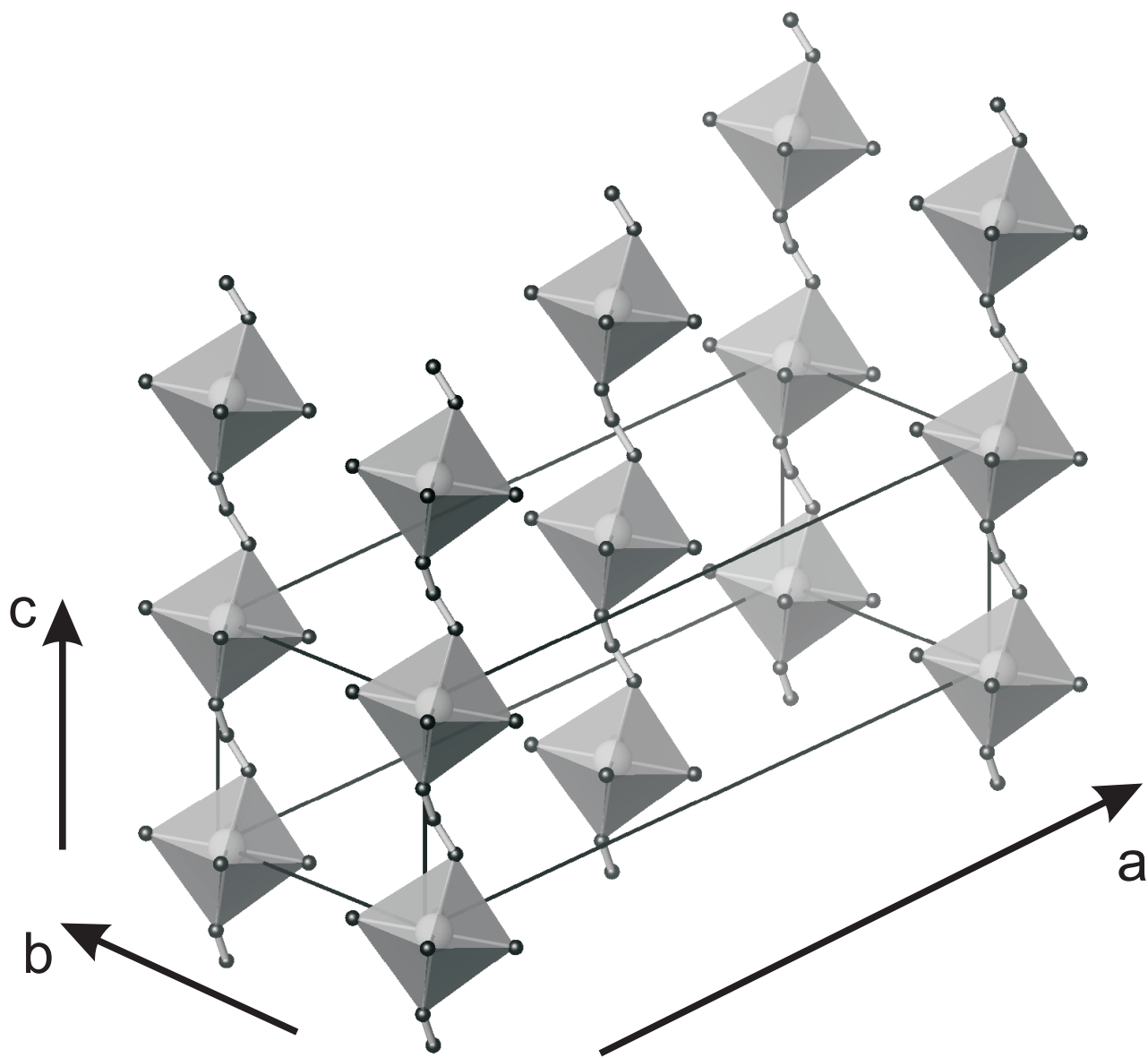
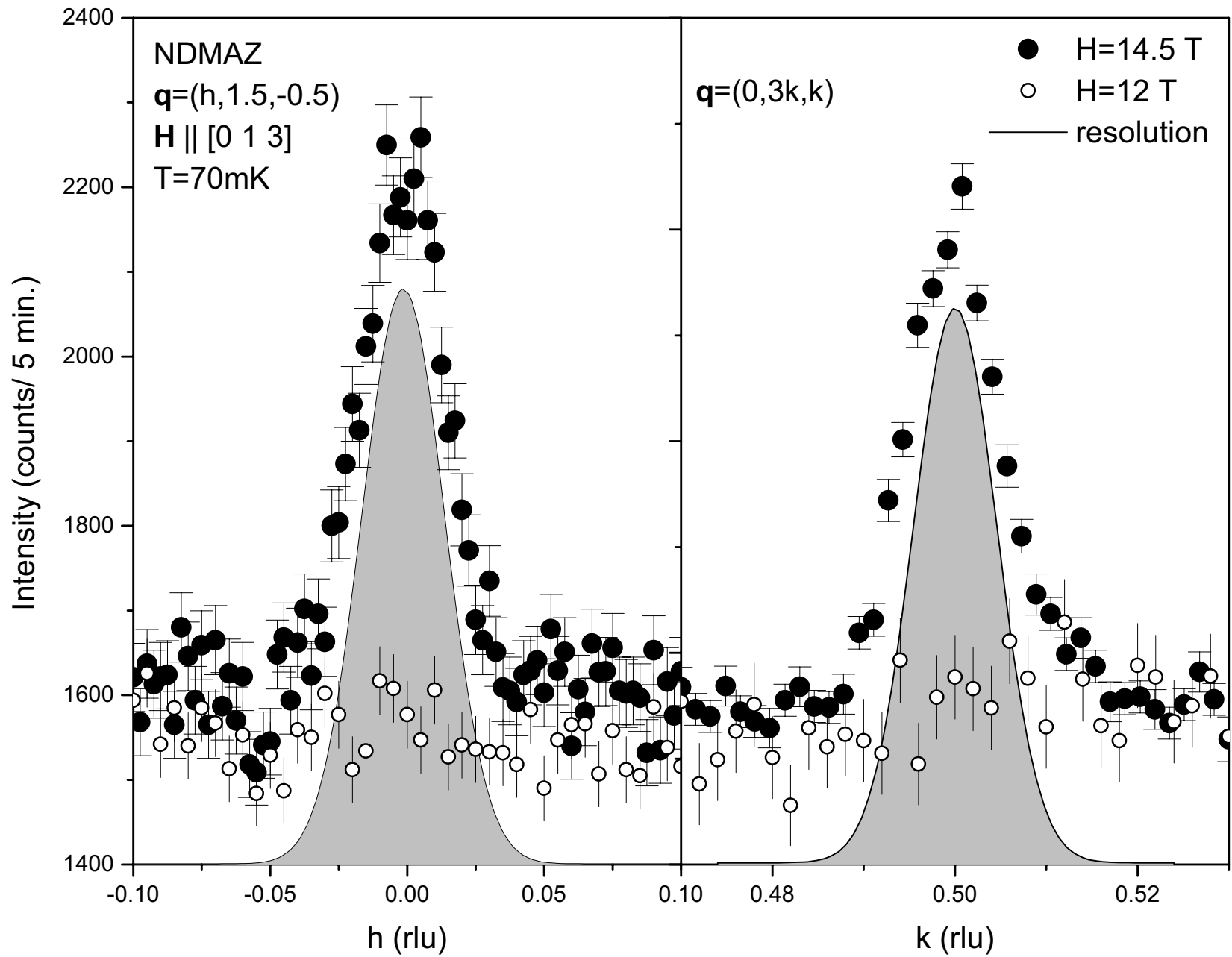
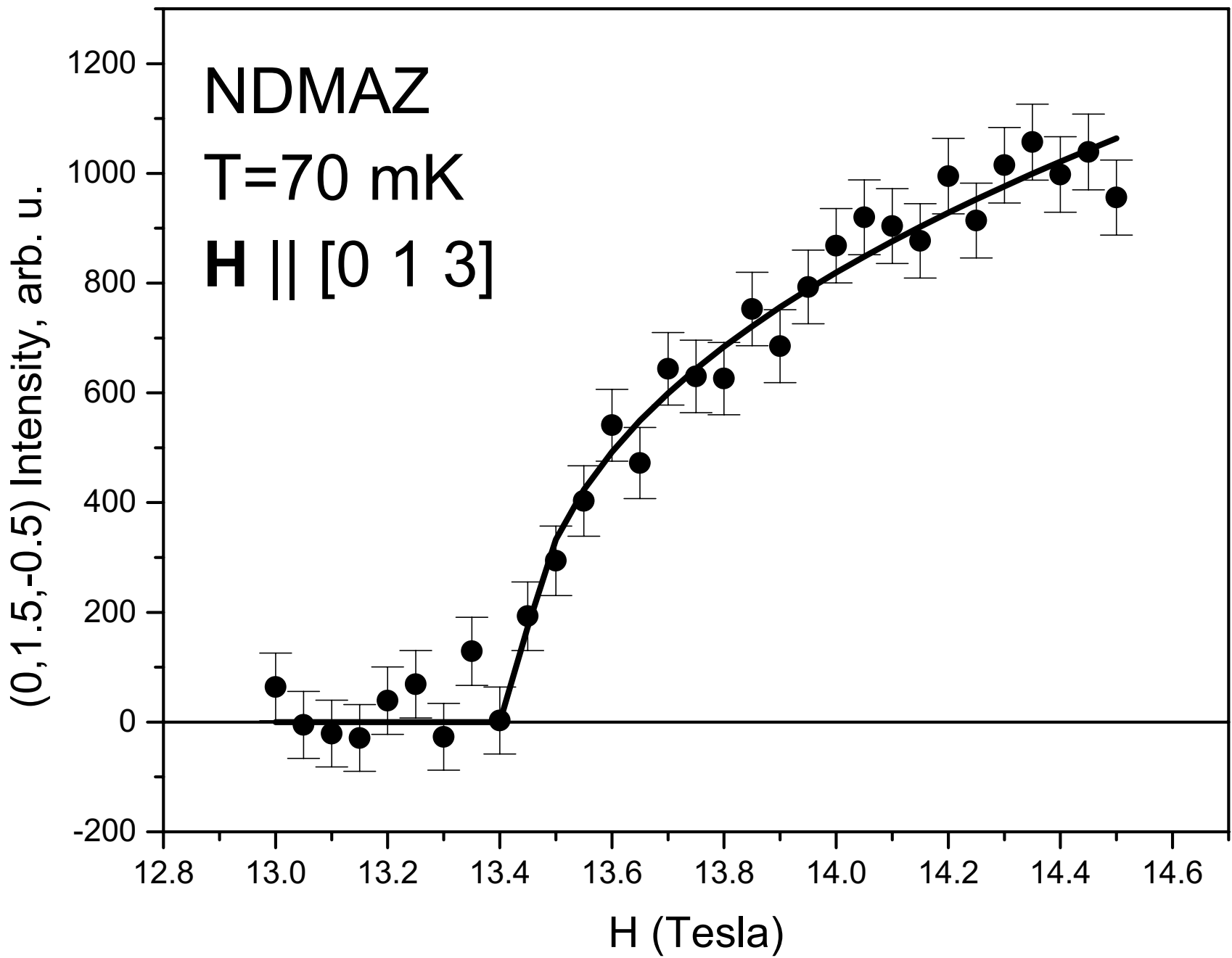
\end{document}